# Less Power for More Learning: Restricting OCaml Features for Effective Teaching

Max Lang[1]          Nico Petzendorfer[1]

[1] Department of Computer Science, Technische Universität München, Garching, Germany
`{max.lang,nico.petzendorfer}@tum.de`

**Abstract.** We present a framework for sandboxing and restricting features of the OCaml programming language to effectively automate the grading of programming exercises, scaling to hundreds of submissions. We describe how to disable language and library features that should not be used to solve a given exercise. We present an overview of an implementation of a mock IO system to allow testing of IO-related exercises in a controlled environment. Finally, we detail a number of security considerations to ensure submitted code remains sandboxed, allowing automatic grading to be trusted without manual verification. The source code of our implementation is publicly available[1].

## Introduction

Programming exercises are a core component of almost any programming course. While it may be feasible to evaluate and grade submissions manually for small numbers of participants, this approach does not scale. Fully automated testing of student submissions is a viable alternative—however, it poses additional technical challenges to ensure grading remains sufficiently accurate.

At the Technical University of Munich, we use OCaml to teach the introductory functional programming course *Functional Programming and Verification*. In large parts, the lecture focuses on *purely* functional programming. While OCaml is certainly well-suited to this paradigm, it also provides rich support for a traditional procedural code style. This poses a problem in a didactic setting: students are drawn to the imperative features of OCaml, which resemble the style of programming they are most comfortable with. Thus, we seek to restrict access to components of OCaml: imperative constructs where they are not desirable, but also arbitrary standard library and syntactical features where appropriate. Automated grading provides an opportunity to both enforce limitations and provide rapid feedback in case they are not followed.

We have developed a framework that allows accurate automated testing of submissions at scale, while enforcing restrictions to the language. Apart from assessing the correctness of submitted code, the framework addresses two key challenges. First, when exercises state restrictions on language and library features, we ensure submissions are bound by those restrictions. Here, the restrictions are made primarily for didactic purposes, and as such misuse is expected to be primarily inadvertent. Second, we ensure that submissions cannot exploit the automated evaluation system, which would make grading unreliable without manual review. Here, we address the small number of students who inevitably attempt to deceive the instructors by maliciously interfering with the evaluation system. Our framework solves these two related problems in such a way that—barring bugs in the implementation—restrictions cannot be circumvented and the automatic grading result can be trusted.

We describe how we have implemented this framework using standard OCaml and Linux tooling, and which considerations are necessary for safely testing untrusted student code at scale. We also describe a mock IO system for effectively testing code making use of OCaml's IO-related features, and other testing components useful in practice. Finally, we report briefly on the use of our framework for teaching over the course of three semesters.

---

[1] As a git repository at https://github.com/just-max/less-power

## Implementation

Before considering the details of how our framework addresses the challenges of automated grading, we provide a high-level overview of the implementation. Our framework uses the Dune build system to build both student code and framework components. Source files from the student's submission are compiled as a Dune library, against which a library implementing the test suite is built. This allows first-class access to the OCaml values of a submission during testing. The test suite is also compiled against a library implementing a known-good sample solution, against which the submission can be compared. Tests are written with QCheck, a property-based testing framework for OCaml, which is inspired by Haskell's QuickCheck library [1]. OUnit provides a test harness around QCheck. The restricted standard library and syntax checker described below are also simply Dune modules. The test framework is run in a containerized environment, with test results output in the common JUnit XML format.

Our framework was developed for use with ArTEMiS, an automated assessment management system for interactive learning [2]. ArTEMiS provides an integrated platform combining exercise problem statements, presentation of feedback, and a continuous integration system. Despite this, our framework is more widely applicable, using only standard tooling, and may easily be adapted to other platforms.

## Automated Grading

Our framework addresses the key challenges of automated evaluation as laid out above. We describe which solutions we have implemented and how they enforce both the didactic restrictions on submissions and the safety features necessary for automated grading.

### Restricted Standard Library and Syntax Elements

The full power of the OCaml standard library, by design, makes many programming tasks easy to solve. For teaching programming, however, students learn most effectively by practicing the concepts they are taught. Therefore, we want to precisely restrict which parts of the OCaml language are available, so that students work with the new concepts instead. This approach of allowing students to focus on learning a limited part of the language is well-known, for instance as *language levels* in *Racket*[2] [3].

*Library features.* For exercises used to teach purely functional programming, standard library features such as references, arrays, and queues should be restricted. For some exercises, we additionally want to restrict standard library features in a more ad-hoc fashion: for example, an exercise where one should implement operations over Peano numbers would be rendered almost trivial with access to OCaml's standard arithmetic operators. Furthermore, some features should be restricted for sandboxing purposes: access to IO would allow a student to give themselves full points on an exercise by simply writing to the appropriate file[3].

The OCaml standard library is modular enough, with limited coupling to the core language, that access to the standard library can easily be restricted in student code. This is a strength of OCaml—similarly restricting standard library features in Java, for example, may be much more involved[4]. Additionally, static guarantees as provided by the OCaml language make such restrictions possible in the first place.

To override the standard library, a new library is defined which contains a module implementing a subset of the interface of `Stdlib`. Since `Stdlib` is plain OCaml source code, this is essentially a copy of `stdlib.ml`—with restricted standard library components removed either directly or using module signatures. Student code is made to use this restricted standard library by compiling with the `-nopervasives` flag to remove all top-level definitions, then using the `-open` flag to instead make our modified standard library available in place of the original. To preclude use of fully qualified references to the standard library (e.g. as `Stdlib.Array`), the new library itself contains a module named `Stdlib` with the same restrictions.

---

[2]Formerly *PLT-Scheme*.
[3]While this is important, it is not security-critical: tests are nevertheless run in a containerized environment.
[4]See, for example, https://github.com/ls1intum/Ares.

*Language-level features.* In addition to the standard library, a number of OCaml's syntactic elements should be restricted[5]. For exercises focused on purely functional programming, loops and mutable records would provide students with procedural programming constructs, instead of requiring them to learn to write functional code. Additionally, regardless of the exact standard library features that are made available, `external` declarations must be disallowed. They would allow students to bypass the restricted standard library by directly accessing OCaml internals. In our framework, the restriction is implemented by iterating the abstract syntax tree of the submission and detecting forbidding syntax elements, an established approach for limiting syntactic features [4].

### Teaching Imperative Concepts

Despite the focus of our teaching on *purely* functional programming, OCaml offers a number of practical impure features which we nevertheless consider highly relevant. As such, we also teach two such imperative features: the IO system of OCaml, and its threading and event system[6].

*IO.* Students should be able to practice their understanding of OCaml's IO system in dedicated programming exercises. However, allowing students direct access to the file system poses problems: along with the safety issues discussed below, using a real file system makes it hard to test that students are properly closing file handles, handling exceptions, and not writing outside of allowed directories.

Our framework makes use of a mock IO system. File handles are implemented as in-memory buffers, along with metadata necessary to simulate a file. The file system hierarchy itself is a simple recursive tree data type. To test the correctness of student code, the state of the mock IO system is reset, the student code called, and finally the state of the mock file system inspected, to check that exactly the correct files were written or created, and that no file handles remain open. The mock IO system maintains handlers for opening a (mocked) file handle from a path, which can be set to raise exceptions upon opening, reading, or writing from the file. This way, it becomes easy to test that submissions properly handle such exceptions. The mock IO system is made available to the student submission in the same way the standard library is restricted.

*Threading and Events.* Additionally, we want students to learn to use OCaml's threading and event system. Unlike with IO, there is no inherent risk posed by simply making these features available. However, we slightly modify the standard library's `Thread` module such that the creation and completion of threads is registered, allowing automated evaluation of requirements such as a maximum number of threads or the correct cleanup of all threads. Unfortunately, terminating rogue threads remains a challenge, and our implementation falls back to process timeouts via OUnit.

### Reliable Automated Testing

Our framework ensures that student code uses only the specified OCaml features and cannot interfere with test results. For teaching at scale, there must be no way to escape this sandbox: it is both unfeasible to manually verify submissions and unrealistic to trust all students to follow the rules. Additionally, the typical security implications of running untrusted code apply.

As part of their OCaml MOOC, Canou et al. propose using the js_of_ocaml compiler to run submitted code in the user's browser [4, 5]. This approach is suitable for providing rapid, high-quality feedback while never executing untrusted code server-side. However, for reliable testing even in a high-stakes situtation, such as exam exercises, we consider malicious attempts to bypass the automated testing restrictions to be a serious issue. Thus, we have found execution in a trusted server-side environment to be necessary.

---

[5]In our default setup, array syntax, `for` and `while` loops, class-related syntax, mutable record fields and `external` declarations are disabled.

[6]OCaml 5's parallelism, being experimental at the time of writing, is not presented.

We can ensure completely reliable automated testing by using the language and library restrictions we have described above. First, the submission is compiled only against our modified standard library. Here, we assume that the mere act of compiling even untrusted OCaml code, is safe. At runtime, student code has access only to the modified standard library it was provided at compile time: here, we are taking advantage of the static guarantees of OCaml. As such, given an appropriate modified standard library, the resulting test executable can be run and the test results trusted.

Beyond reliable automatic grading, a number of safety issues are addressed. A key objective is to prevent sensitive files from within the sandbox from being leaked to the student. Making IO unavailable in the modified standard library takes care of this at runtime. Additionally, while compiler messages are normally helpful, if produced while compiling the test or sample solution library, they may leak details of the sample solution. Thus, standard error output of the compiler is dropped for all modules except the submission.

### Testing and Evaluating

Once restrictions, mocks, and safety features have been considered, the task remains of evaluating student submissions for correctness. To make it possible to create new exercises and iterate on existing exercises with a reasonable amount of effort, we use the QCheck library for property-based testing. QCheck provides combinators for generating inputs to student code, for *shrinking* the generated inputs to smaller inputs that make feedback easier to understand, and for creating runnable tests. The library OUnit integrates with QCheck to execute tests and provide output in a suitable format. We have found the QCheck library to be effective for writing high-quality test cases with a reasonable amount of work.

Finally, we make use of some OCaml platform tools for evaluating some more specialized exercises. Using the stack size limit parameter to the OCaml bytecode runtime, we obtain a practical method to check that tail recursion has been used to avoid stack overflow. Additionally, for testing exercises where students are expected to write code from scratch, without a type-correct skeleton, we create a series of mini-libraries. Each library compiles successfully only if an individual value from the student submission is type-correct. Then using Cppo, a C-style preprocessor, a compile-time decision is made in the test library to reference either the value from the submission or a known type-correct dummy implementation.

### Usage Report

Using the presented framework, we have successfully run the course *Functional Programming and Verification* over three semesters. Each semester has included seven weeks of OCaml programming exercises, as well as two separate exams, with between 500 and 1000 participating students. From monitoring random samples of student submissions, we have found no successful attempt to bypass the restrictions we have imposed. The framework, when paired with the ArTEMiS submission system, has made it possible to offer nearly immediate feedback on programming exercises, even in exams.